\documentclass[manuscript]{aastex}

\newcommand\rah{\mbox{$^{\mathrm h}$}}%
\newcommand\ram{\mbox{$^{\mathrm m}$}}%

\slugcomment{To appear in AJ}

\shorttitle{Parsec-scale shocks in the kiloparsec-scale jet of Centaurus A}
\shortauthors{Tingay \& Lenc}

\begin{document}


\title{Parsec-scale shocks in the kiloparsec-scale jet of Centaurus A}


\author{S.J. Tingay}
\affil{Curtin Institute of Radio Astronomy, Curtin University of Technology, Bentley WA, Australia}
\email{s.tingay@curtin.edu.au}

\author{E. Lenc}
\affil{CSIRO Australia Telescope National Facility, P.O. Box 76, Epping, NSW 2121, Australia}



\begin{abstract}
High angular resolution VLBI observations of Centaurus A have been undertaken that allow access to a wide field-of-view, encompassing both the well-studied pc-scale jet and the inner part of the kpc-scale jet.  The VLBI observations have detected compact regions of synchrotron emission in the kpc-scale jet that coincide with three stationary features identified from previous VLA monitoring observations.  Each of these stationary features is associated with strong localised X-ray emission.  The VLBI results strengthen arguments made by previous authors suggesting that the stationary features may be the result of stellar objects or gas clouds traversing the jet flow, intercepting the jet and causing strong shocks.  The VLBI data show that the most strongly shocked regions in these features are resolved but have extents no larger than a few pc, reducing the required mass of the typical intercepting object by a factor of $\sim$10 relative to previous estimates, making explanations based on high mass loss stars or low density gas clouds more plausible.
\end{abstract}


\section{Introduction}

Centaurus A, the Fanaroff-Riley type I radio galaxy hosted in the nearby galaxy NGC 5128, is the closest classical radio galaxy to us and has been the target of a number of high angular resolution investigations with a variety of instruments operating across the electromagnetic spectrum.  For a general review, see \cite{is98}.  As a nearby radio galaxy (3.84$\pm$0.35 Mpc; \cite{rej04}), high resolution studies of the active nucleus of the galaxy are of special interest, in order to probe as closely as possible to the $5.5 \times 10^{7}$ solar mass black hole \citep{cap09}, with the aim of gaining insight into the processes giving rise to relativistic jets (1 mas $\sim$ 0.02 pc at 3.84 Mpc).

The active nucleus of Centaurus A has been the subject of comprehensive very long baseline interferometry (VLBI) observations, the highest angular resolution direct imaging technique in astronomy.  In a series of studies, the structure and evolution of the sub-parsec-scale jet originating at the galaxy nucleus were explored \citep{hor06, tin01a, tin01b, fuj00, tin98,  vanl98, kel97, jon96}.  These probes of the relativistic jet material close to its source have yielded information on the speed of flow of the material, the bipolar nature of the flow (a jet and counterjet), the size of the ``core'' of the radio source relative to the expected Schwarschild radius of the black hole, and the nature of the ionised and molecular environment of the active nucleus.

In recent years the VLBI technique has also been used to help determine the nature of the jets originating from active nuclei on the vastly larger kiloparsec scale.  VLBI can be used to detect compact radio emission from shocks in radio galaxy jets after they have propagated large distances from the active nucleus.  Recently, \cite{tin08} have used VLBI observations to detect the compact shocked structures in the termination hot spot of the closest Fanaroff-Riley type II radio galaxy, Pictor A, providing the highest resolution view of a radio galaxy hot spot to date and providing new empirical evidence that leads to a more natural interpretation of the X-rays from the Pictor A hot spot than previously possible.  \cite{tin08} also review other examples of VLBI observations of radio galaxy termination hot spots.

In between the active nucleus and the termination hot spots in radio galaxies, the jet propagates large distances and it is supposed that brightness enhancements in the jets seen at arcsecond-scale resolution with instruments like the Very Large Array (VLA) are locally shocked regions, due to internal instabilities in the flow of material or due to interactions with ambient material external to the jet.  These enhancements are typically of much lower brightness temperature than the radio emission observed from either the active nucleus or the termination hot spots.  However, a sensitive VLBI search for compact structure in these shocked regions may be of great use in understanding the propagation of the jets and the processes by which they are shocked.

Here we describe such a search for compact structures in the kiloparsec-scale jet of Centaurus A.  The jet on these scales has been very well studied with the VLA over a long period of time.  Most recently, \cite{har03} combined a multi-epoch study of the kiloparsec-scale jet of Centaurus A with the VLA and new {\it Chandra} X-ray data.  The proximity of Centaurus A is crucial in this study, as Centaurus A is the only radio galaxy for which the {\it Chandra} angular resolution is comparable to the energy-loss travel distance of the X-ray emitting electrons.  Centaurus A is therefore an important case study for understanding jets from active galaxies and the \cite{har03} results reveal much about the dynamics of the jet on the kiloparsec-scale, based on radio and X-ray observations of the shocked regions of the jet.  VLBI observations can add further information, by probing the conditions within the shocked regions at high resolution.

In section 2 we present our VLBI observational data and imaging results.  In section 3 we discuss the implications of our detection of compact structures in the shocked regions of the kiloparsec-scale jet, relating them to the previous results of \cite{har03}.

\section{Observations and data analysis}
\subsection{LBA Observations}
A VLBI observation of Centaurus A (PKS B1322$-$428) was made on 21 June, 2007 using a number of the Australian Long Baseline Array (LBA) telescopes: the 70 m NASA Deep Space Network (DSN) antenna at Tidbinbilla (5 hr track, RR polarisation only); the 64 m antenna of the Australia Telescope National Facility (ATNF) near Parkes (10 hr track); 5 $\times$ 22 m antennas of the ATNF Australia Telescope Compact Array (ATCA) near Narrabri used as a phased array (12 hr track); the ATNF Mopra 22 m antenna near Coonabarabran (10 hr track); the University of Tasmania's 26 m antenna near Hobart (9 hr track); and the University of Tasmania's 30 m antenna near Ceduna (11 hr track).  The observation utilised hard disk data recorders (Phillips et al. 2009, in preparation); allowing dual circular polarisation data across $4\times16$ MHz bands (2268-2332 MHz) to be recorded at Parkes, ATCA, and Mopra, and $2\times16$ MHz bands (2268-2300 MHz) at the remaining antennas. Observing scans were centred on the nucleus of Centaurus A ($\alpha = 13\rah25\ram27\fs615200$; $\delta = -43\arcdeg1\arcmin8\farcs80500$ [J2000]) which doubled as a phase reference source.

The data were correlated with the DiFX Software Correlator \citep{Deller:2007p10545} using an integration time of 2 seconds and 64 frequency channels across each 16 MHz band (channel widths of 0.25 MHz). The ATCA primary beam limits the field of view of this observation to a half width at half maximum (HWHM) of $\sim47\arcsec$. At the HWHM point, bandwidth smearing (a form of chromatic aberration) and time-averaging smearing \citep{Cotton:1999p19452,Bridle:1999p10564} losses are estimated to be approximately 5\% and 4\%, respectively.

\subsection{LBA Data Reduction}

The correlated data were imported into the AIPS\footnote{The Astronomical Image Processing System (AIPS) was developed and is maintained by the National Radio Astronomy Observatory, which is operated by Associated Universities, Inc., under co-operative agreement with the National Science Foundation} package for initial processing. The data were fringe-fit (AIPS task FRING) using a one minute solution interval, finding independent solutions for each of the 16 MHz bands.

\begin{figure}
\epsscale{1.0}
\plotone{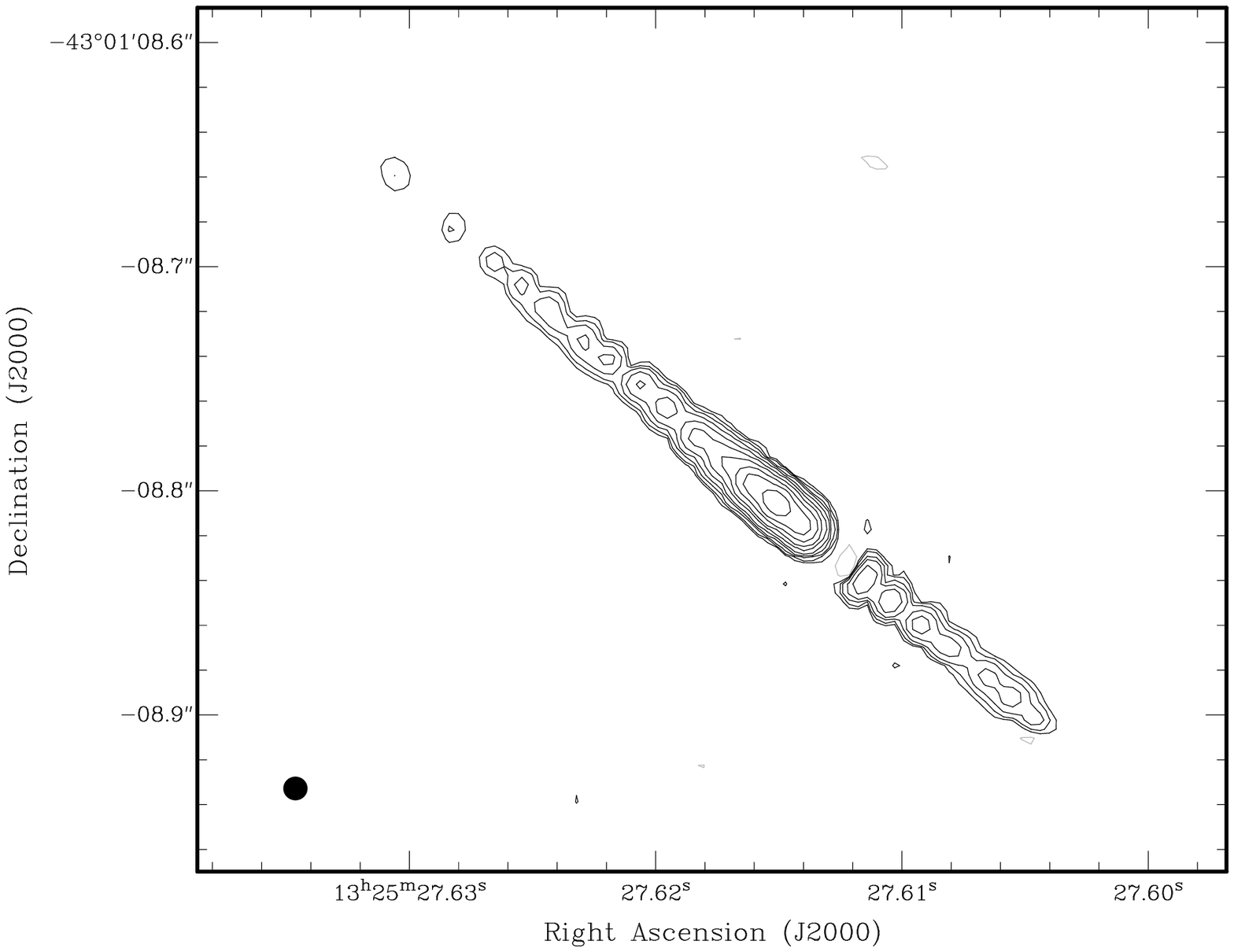}
\caption{Uniformly-weighted 2.3 GHz LBA image of the jet and counterjet in Cen A. The peak surface brightness is 1.46 Jy beam$^{-1}$ and the $1\sigma$ rms image noise is 0.27 mJy beam$^{-1}$. Contours are drawn at $\pm1, \pm2, \pm4, \pm8, \cdots$ times the $6\sigma$ rms noise. The beam size is $10.8\times10.5$ mas at a position angle of $37\arcdeg$.}
\label{fig:figjet}            
\end{figure}

The nominal system temperatures applied for each antenna during correlation were refined by applying the antenna system temperatures measured throughout the observation. Further refinement in the calibration was complicated by the complex jet structure of Centaurus A. To account for this structure a new DIFMAP \citep{Shepherd:1994p10583} task, \emph{cordump}\footnote{The \emph{cordump} patch is available for DIFMAP at \url{http://astronomy.swin.edu.au/$\sim$elenc/DifmapPatches/}} \citep{Lenc:2006p32}, was used to transfer all phase and amplitude corrections made in DIFMAP during the imaging process to an AIPS compatible solution table. The Centaurus A data were averaged in frequency and exported to DIFMAP where several iterations of modelling and self-calibration of both phases and amplitudes were performed. The resulting contour map of the jet and counter-jet are shown in Figure \ref{fig:figjet}. A one sigma RMS noise of 0.27 mJy beam$^{-1}$ is measured in the final image; just over twice the theoretical thermal noise for the observation\footnote{Estimated with the ATNF VLBI sensitivity calculator: \url{http://www.atnf.csiro.au/vlbi/calculator/}}. The phase and amplitude corrections were transferred back to the unaveraged AIPS data set with \emph{cordump} and the jet model components were subtracted from the calibrated data set (AIPS task UVSUB) to simplify detection of sources in the wider field.

\begin{figure}
\epsscale{1.0}
\plotone{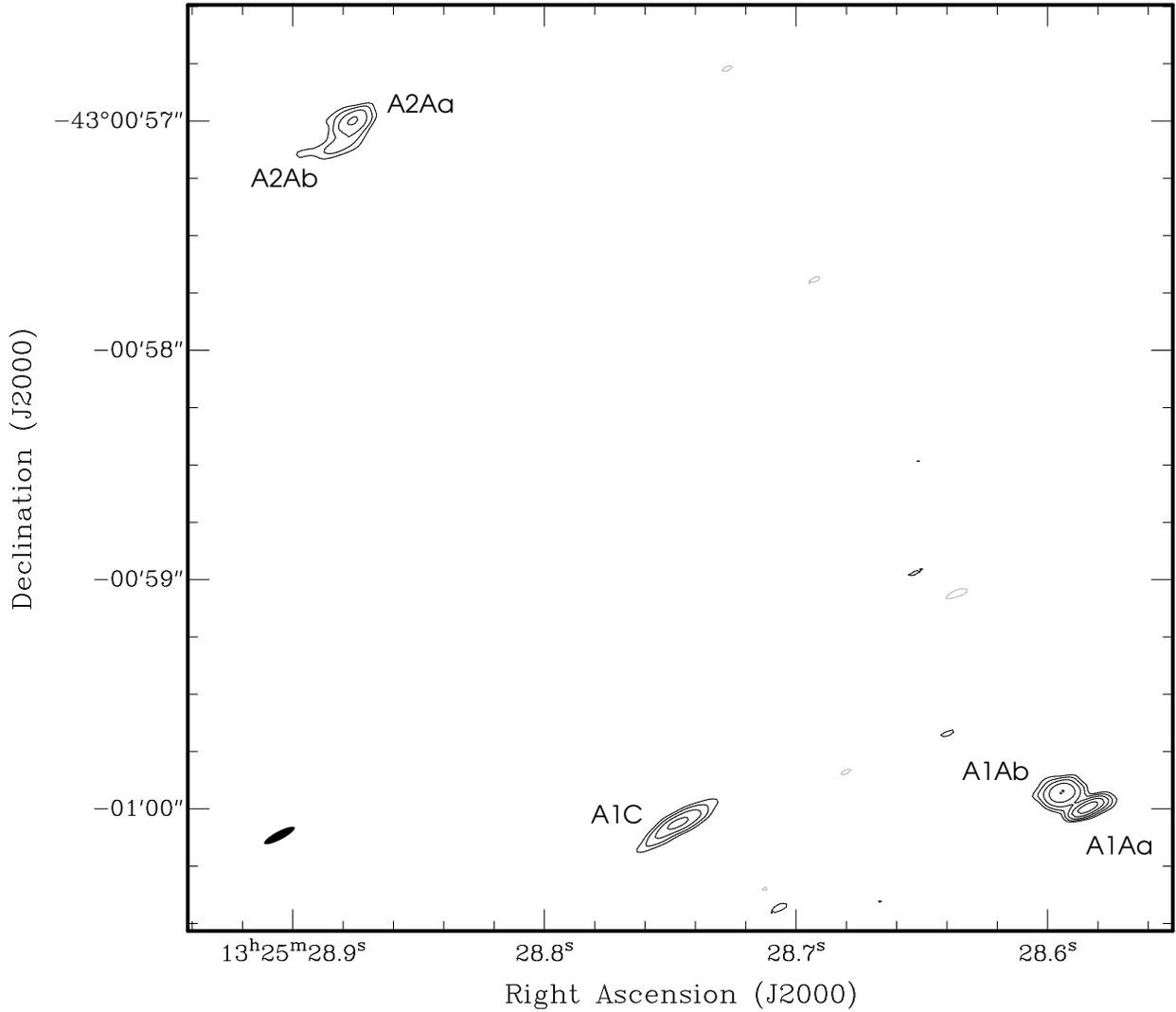}
\caption{Naturally-weighted 2.3 GHz LBA image of knot features in Cen A. The peak surface brightness is 3.8 mJy beam$^{-1}$ and the $1\sigma$ rms image noise is 0.25 mJy beam$^{-1}$. Contours are drawn at $\pm2^{0}, \pm2^{\frac{1}{2}}, \pm2^{1}, \pm2^{\frac{3}{2}}, \cdots$ times the $3\sigma$ rms noise. The beam size is $150\times37$ mas at a position angle of $-62\arcdeg$.}
\label{fig:figknots}            
\end{figure}

To reduce non-coplanar effects \citep{Perley:1999p10576}, the visibilities were $(u,v)$ shifted (AIPS task UVFIX) approximately $16\arcsec$ north-east of the nucleus so that the new phase center would coincide with bright knots observed with the VLA \citep{har03}. Initial imaging of the knot region, using all baselines and uniform weighting, revealed no detections above a $6\sigma$ detection threshold suggesting that sources in this region were either extended, weak, or both. A second imaging attempt included only the shortest baselines (between Parkes, ATCA and Mopra) and natural weighting to improve sensitivity to larger scale structure. The lower resolution image revealed four detections above the $6\sigma$ threshold (Figure \ref{fig:figknots}) and achieved a one sigma rms noise of 0.25 mJy beam$^{-1}$. All images were produced for display using the KARMA software package \citep{Gooch:1996p7263}.

\section{Results}
\subsection{Identification of Sources}
\label{sec:sourceid}
A list of sources detected above the $6\sigma$ detection threshold of the short-baseline VLBI data is given in Table \ref{tab:tabfluxes}. The $6\sigma$ threshold provides a conservative false detection rate of less than one in 3000. Flux density errors of $\pm10\%$ are expected due to uncertainties in the absolute flux density scale for Southern Hemisphere VLBI \citep{rey94}. The total flux density and sizes of each source were determined by modelling each component with elliptical Gaussian components, the peak flux density, integrated flux density, fitted size and position angle (PA) of each source is listed in Table \ref{tab:tabfluxes}.

\begin{table}[ht]
\begin{center}
\begin{tabular}{lccccc} \hline \hline
Component & Size           & P.A.\tablenotemark{a}  & $S_{2.3}$ Peak    & $S_{2.3}$ Integrated & $T_{B}$ \\
          & (mas)          & (deg)                  & (mJy beam$^{-1}$) & (mJy)                & (K)     \\ \hline \hline
A1Aa      & $83\times39$   & -73                    & $3.8$               & $6.1$                  & $4.3\times10^{5}$ \\
A1Ab      & $109\times45$  & 47                     & $3.1$               & $9.5$                  & $4.5\times10^{5}$ \\
A1C       & $260\times70$  & -59                    & $2.4$               & $9.8$                  & $1.2\times10^{5}$ \\
A2Aa      & $106\times106$ & -40                    & $2.2$               & $8.0$                  & $1.6\times10^{5}$ \\
A2Ab      & $320\times64$  & -83                    & $1.5$               & $5.6$                  & $6.3\times10^{4}$ \\ \hline
\tablenotetext{a}{Position angle of major axis measured east of north.}
\end{tabular}
\caption{VLBI component parameters.}
\label{tab:tabfluxes}
\end{center}
\end{table}

Figure 2 shows the detection of the weak but compact radio sources listed in Table 1.  An overlay of the \cite{har03} image and our VLBI image (Figure 3) shows that the weak, compact radio sources coincide with features A1A (resolved into subcomponents we designate A1Aa and A1Ab), A1C and A2A (resolved into subcomponents we designate A2Aa and A2Ab).  These compact sources are each better than seven times the RMS noise level in the VLBI image of 0.25 mJy/beam and are therefore detected with good confidence. On the basis of the VLA flux densities presented in \cite{har03}, one other component may have been expected to be detected, A1B.  A1B was not detected in our VLBI data.

\section{Discussion}

\cite{har03} examine a number of different possibilities for the nature of the discrete features in the jet of Centaurus A, detected at both radio and X-ray wavelengths.  We examine the new VLBI imaging results in the light of these possibilities. 

\cite{har03} briefly consider the possibility that the features are jet-associated radio supernovae or supernova remnants.  Certainly at the distance of Centaurus A, the compact objects we have detected with VLBI, at a few mJy each, are typical (in terms of flux density) of radio supernova remnants detected in nearby starburst galaxies such as M82 \citep{mac01}, NGC 253 \citep{len06, tin04} or NGC 4945 \citep{len08}.  While very few supernova remnants seen in other galaxies can be resolved into a shell morphology, some clearly are.  The fact that the compact sources detected in Figure 2 do not exhibit shell morphologies is not necessarily evidence against them as supernova remnants.  The evidence against these sources being supernova remnants, as argued by \cite{har03}, is that the X-ray emission associated with these radio components is apparently non-thermal.

\cite{har03} also argue against the possibility that the jet features are purely compressions in the synchrotron-emitting fluid, suggesting that the features instead represent privileged sites of in situ particle acceleration.  These sites may be standing shocks in the fluid flow, caused by the jet's interaction with a stationary object of some description - high mass loss stars or gas clouds are cited as possible examples.  \cite{har03} suggest that the A1 complex of components may represent standing shocks related to the transition from the collimated inner jet to the flared extended jet, reserving the interpretation of jet interaction with stationary objects for the stationary components (stationary simply meaning that no proper motion has been detected) downstream of the A1 complex (for example component A2A).  However, it is apparent from Figure 1 in \cite{har03} that the inner jet commences flaring well before the jet reaches the A1 complex.  Why components A1A and A1C should therefore be stationary shocks at that particular point in the jet is not clear.  Under this interpretation it is also not obvious why component A1B should have a large proper motion as part of the complex of shocks at the jet transition point.  A common interpretation for all of the stationary components in the jet therefore remains a possibility, that all stationary radio features are related to interactions with objects that have intercepted the jet flow.

From our VLBI observations we find that we detect compact radio emission from the VLA features A1A, A1C and A2A.  From the \cite{har03} {\it Chandra} data, these features correspond to AX1A, AX1C and AX2.  These are all significant X-ray sources and all have radio counterparts in the VLA image that are stationary.  The VLBI data now show that these three features have radio emission compact on parsec scales.  Simply on the basis of the VLA radio flux densities as a predictor of compact emission on VLBI scales, one other compact radio component may have been expected to be detected with VLBI, along with A1A, A1C and A2A, that is A1B.  A1B was not detected with VLBI, although it is of comparable strength and size to the other three on VLA scales.  A1B is the only feature of the four that has a large apparent motion away from the nucleus.  The proposition of \cite{har03} that the stationary features in the Centaurus A jet represent stationary shocks is strengthened by our VLBI results that show compact radio structures associated with these stationary features.

The sizes of the compact radio sources detected with VLBI are all less than $\sim$6 pc (typical size of $\sim$2 - 3 pc), significantly less than the 10 pc assumed by \cite{har03} when calculating the properties of the objects that could possibly explain the presence of the stationary shocks.  The sizes of the sub-structures seen in the A1Aa and A1Ab components at VLBI resolution are consistent with the sizes inferred from $\sim$10\% flux density variability over a decade detected in the VLA data \citep{har03}.  We note that due to sensitivity and surface brightness sensitivity limitations with the VLBI data, weaker and/or more extended emission could be associated with the compact components, the structure in the VLBI images perhaps representing the core of the shocked region.  However, the primary interactions in the shocked regions are likely to be of order only a few pc in extent.  Taking the VLBI sizes for the shocked regions as indicative of the sizes of the interaction regions in the jet implies that the constraint on the required mass in the obstacles is reduced by a factor of $\sim$10, compared to estimates based on the component sizes as measured with the VLA (see equation in section 5.3 of \cite{har03} that has the mass of the obstacle proportional to the square of the radius of the obstacle).

This reduction in the required mass of the obstacle, from at least a few solar masses to substantially less than one solar mass, relaxes some of the extreme characteristics required by \cite{har03} when discussing obstacles based on high mass loss stars and low density gas clouds

The possibility that the stationary shocked regions could be due to gas clouds intercepting the jet is a scenario that has been studied by a number of authors using numerical simulations.  Recently \cite{jey09} reported hydrodynamical simulations of jets interacting with clouds of various sizes that move through the jet at various speeds, showing the amount of disruption caused.  Judged qualitatively against the images showing the kpc-scale morphology of the Centaurus A jets, one could conclude that the stationary features in the kpc-scale jet are due to relatively fast moving clouds ($\sim$10,000 km s$^{-1}$) with diameters only a small fraction of the jet width.  The VLBI results can be taken to support such a conclusion - a cloud diameter of a few pc covering approximately 5\% of the jet width.

A comprehensive recent study of jet interactions with stationary clouds by \cite{cho07} shows that compact synchrotron emission can be generated from these types of interactions.  Again, qualitatively, the compact synchrotron-emitting structures seen in the numerical simulations appear similar to those observed with VLBI, in particular components A1Aa/b and A2Aa/b in Figure 2 that show a bright compact feature and a weaker and more extended trailing feature.

\section{Conclusions}

We have made the first VLBI images of pc-scale structures in the kpc-scale jet of Centuarus A, detecting compact regions of synchrotron emission.  All detected pc-scale components coincide with the locations of jet components seen to be stationary from previous VLA monitoring of other authors.  Other features in the kpc-scale jet have been shown to have large proper motions, from the VLA monitoring.  None of these moving components have been detected with VLBI.  These results support suggestions that the stationary components in the jet may be due to moving stellar objects or gas clouds intersecting the jet flow and causing strong shocks.  Size estimates for the shocked regions suggest extents of order a few pc, smaller than previously assumed and consistent with observed flux density variability, leading to approximately an order of magnitude reduction in the required mass for the obstacles.  This reduction relaxes some of the extreme constraints on the inferred properties of high mass loss stars and low density gas clouds suggested as providing plausible obstacles.  A repeat of these VLBI observations at a later date will put limits on proper motions for the detected pc-scale components in the kpc-scale jet of Centaurus A, allowing further study of their properties.

\acknowledgments
This research has made use of NASA's Astrophysics Data System.  This research has made use of the NASA/IPAC Extragalactic Database (NED) which is operated by the Jet Propulsion Laboratory, California Institute of Technology, under contract with the National Aeronautics and Space Administration.  SJT acknowledges the support of the Western Australian State government, in the form of a Western Australia Premier's Fellowship.

{\it Facilities:} \facility{LBA}.

\clearpage

\begin{figure}
\epsscale{1.0}
\plotone{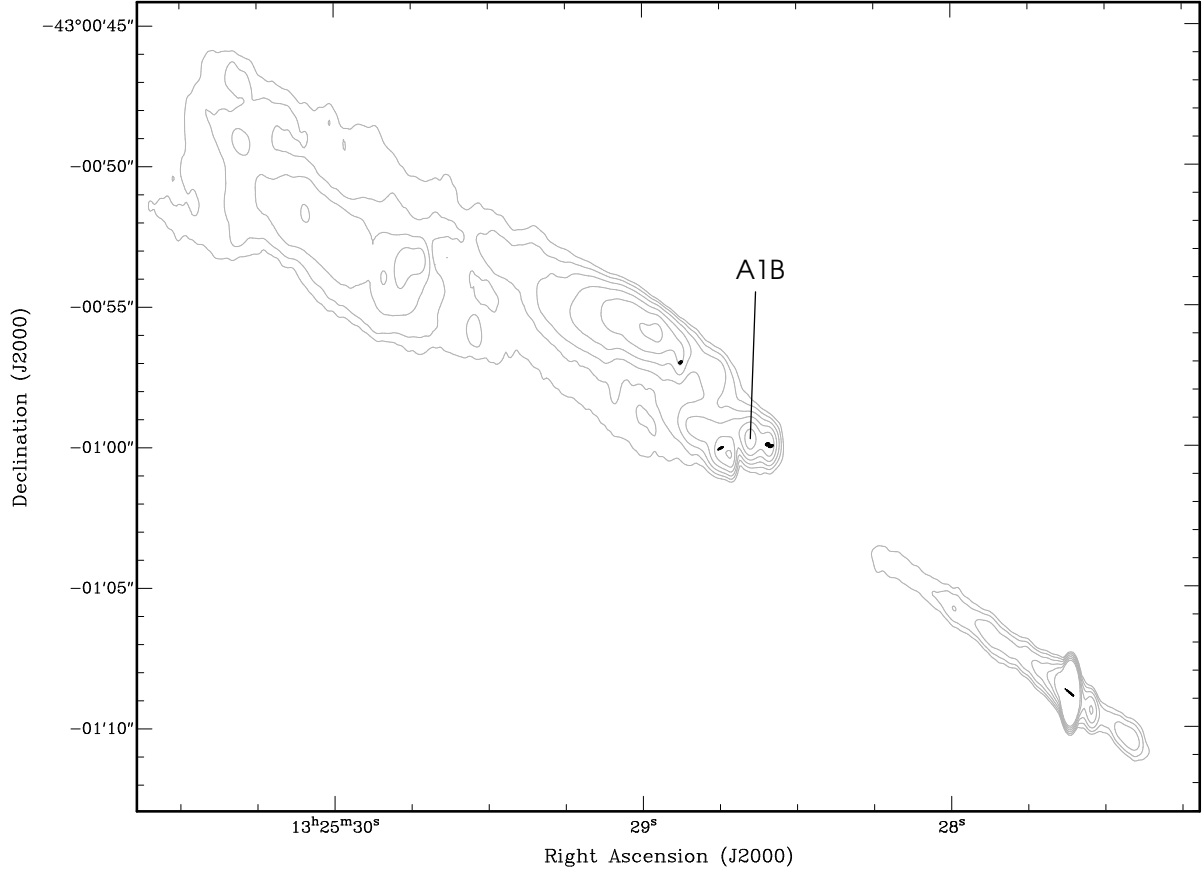}
\caption{Uniformly-weighted 8.4 GHz VLA contour map of Centaurus A (grey contours) overlaid with a naturally-weighted 2.3 GHz LBA contour map (black contours) to show the locations of the compact sources detected within the jet and knot features.}
\label{fig:figovl}            
\end{figure}

\end{document}